\documentclass[aps,twocolumn,pra,showpacs,superscriptaddress]{revtex4}
\usepackage{graphicx}

\newcommand{\be}{\begin{equation}}
\newcommand{\ee}{\end{equation}}
\newcommand{\bea}{\begin{eqnarray}}
\newcommand{\eea}{\end{eqnarray}}

\begin{document}

\title{Single-atom detection using whispering gallery modes of microdisk
resonators}

\author{Michael Rosenblit}
\affiliation{Department of Physics and Ilse Katz Center for Meso-
and Nanoscale Science and Technology, Ben Gurion University of the
Negev, P.O.Box 653, Be'er Sheva 84105, Israel}

\author{Peter Horak}
\affiliation{Optoelectronics Research Centre, University of
Southampton, Southampton SO17 1BJ, United Kingdom}

\author{Steve Helsby}
\affiliation{Optoelectronics Research Centre, University of
Southampton, Southampton SO17 1BJ, United Kingdom}

\author{Ron Folman}
\affiliation{Department of Physics and Ilse Katz Center for Meso-
and Nanoscale Science and Technology, Ben Gurion University of the
Negev, P.O.Box 653, Be'er Sheva 84105, Israel}

\date{\today}

\pacs{42.50.Ct, 42.82.-m, 32.80.Qk}

\begin{abstract}
We investigate the possibility of using dielectric microdisk
resonators for the optical detection of single atoms trapped and
cooled in magnetic microtraps near the surface of a substrate. The
bound and evanescent fields of optical whispering gallery modes
are calculated and the coupling to straight waveguides is
investigated using finite-difference time domain solutions of
Maxwell's equations. Results are compared with semi-analytical
solutions based on coupled mode theory. We discuss atom detection
efficiencies and the feasibility of non-destructive measurements
in such a system depending on key parameters such as disk size,
disk-waveguide coupling, and scattering losses.
\end{abstract}

\maketitle

\section{Introduction}

Micro resonators are a promising system for research in a wide
range of fields. Their spectral properties can be exploited in
various applications ranging from telecommunication \cite{slusher}
to biological/chemical sensors \cite{vollmer} as well as in
fundamental research. Of specific interest is the potential
contribution of such devices to the emerging field of quantum
technology (QT), which may again serve as an enabling technology
for both fundamental science and applicative research.

In the realm of QT, single atoms and ions coupled to micro
resonators are one of the most promising systems. There, a high
degree of control over light-atom interaction can be achieved,
which may enable new insight and capabilities in contexts such as
cavity QED \cite{QED}, single photon sources \cite{pelton},
memory, and purifiers for quantum communication \cite{qipc},
manipulation of internal and external degrees of freedom for
matter waves in systems like interferometric sensors
\cite{andersson}, clocks, and perhaps also the elusive quantum
computer \cite{computer,qipc}. In this paper we will analyze a
specific micro resonator in the form of a microdisk, and
furthermore we will focus on its ability to detect single atoms.

Recently reported progress in the manufacturing of high Q
dielectric microdisk structures \cite{Kippenberg} motivates the
development of compact and integrable devices. Of specific
interest is the wafer-based manufacturing of resonators where a
good control of the physical characteristics can be achieved
during fabrication enabling, for example, extremely small mode
volumes as well as accurate alignment with other elements such as
microtraps. Furthermore, a wafer based device may allow more
complex functions such as tunability to be integrated.

Experiments have also shown that it is possible to trap, guide and
manipulate cold, neutral atoms in miniaturized magnetic traps
above a substrate using either microscopic patterns of permanent
magnetization in a film or microfabricated wire structures
carrying current or charge \cite{folman,reichel}. Such surfaces
have received the name atom chip.

Recently it was proposed to integrate micro-optics into an atom
chip for atom-light interaction. In particular, the use of
Fabry-Perot \cite{horak} or photonic bandgap \cite{lev} cavities
has been discussed. As a complementary and comparative analysis we
investigate in this work the external fields of optical whispering
gallery modes which are supported by a toroidal microcavity and
which may be used for the above purpose. Such a microdisk can be
considered as a two-dimensional version of the much studied
microsphere \cite{lefevre,klitzing,kimble}.

This work is organized as follows: In Sec.\ \ref{sec:model} we
describe the system under consideration. Section \ref{sec:wgm}
deals with the different numerical and analytical methods which we
apply to examine the disk resonator structures. We analyze
feasible realizations, taking into account imperfections of
various origins. We then introduce atoms to the system in the
framework of the Jaynes-Cummings model (Sec.\ \ref{sec:coupling}).
Subsequently, the calculated fields are used to estimate the
efficiency of our scheme for single atom detection (Sec.\
\ref{sec:detection}). In Sec.\ \ref{sec:experiment} we discuss
experimental considerations such as tunablity. Finally, we
conclude in Sec.\ \ref{sec:conclusions}.


\section{System design}
\label{sec:model}

\begin{figure}
\includegraphics[height=6.5cm]{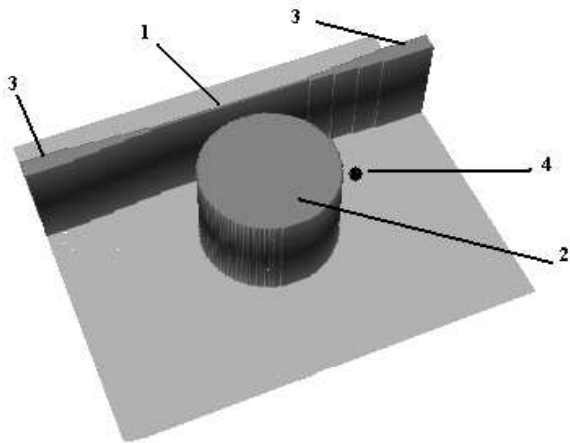}
\caption{The structure under consideration. The evanescent wave
from the slab waveguide (1) is coupled into the disk (2) and back
through a small gap between them. The adiabatic waveguide tapers
(3) serve for coupling light from optical fibers (not shown) into
and out of the waveguide. Cold atoms (4) can be positioned on the
disk side. \label{fig:structure}}
\end{figure}

The basic system under discussion is that of an atom chip
consisting of a magnetic or dipole microtrap for cold atoms, and a
photonic part which is used for optical atom detection. As shown
in Fig.\ \ref{fig:structure}, in this work we focus on a photonic
system consisting of a circular plate (disk) and a linear
waveguide.

The waveguide couples light into and out of the disk. In a
realistic setup, both waveguide ends will be attached to optical
fibers. In order to optimize power transfer between the waveguide
and a single-mode fiber, the mode overlap at the interface has to
be maximized. This requires waveguide dimensions of 9x9 to 12x12
microns in the refractive index range of 1.454 - 2.17 for
wavelengths around 780nm (the wavelength of Rb atoms usually used
in this kind of atom chip experiments). In this case, best
coupling efficiencies of the order of 96-98\% can be achieved. For
best mode matching with the microdisk modes, the waveguide width
has to be reduced to about 0.3 to 1.2$\mu$m using adiabatic
tapers, as shown in Fig.\ \ref{fig:structure}.

The disk itself supports a large range of resonant modes. Here, we
are mainly interested in the low-loss modes traveling along the
disk edge in the form of whispering gallery modes (WGM). While
most of the mode energy is confined within the disk, a small part
of the mode exists outside the disk as an evanescent field, and it
is here that an atom can interact with the light. This atom-light
coupling changes the optical properties of the disk mode, which
subsequently changes the intensity and/or phase of the light at
the output of the linear waveguide. These changes can then be
measured to infer the presence of the atom.

In the horizontal direction the linear waveguide and the disk are
bordered by air or vacuum with refractive index $n_{cl} = 1$. In
our calculations we assumed structures made of fused silica with
$n_{c}=1.454$ at a wavelength of 780nm. In the vertical direction
the structure may be more complicated with several layers in order
to give good mode confinement.

A number of numerical and analytical tools have been utilized to
investigate the optical properties of such systems, for example
coupled mode theory (CMT) \cite{hammer,love}, scattering matrix
theory \cite{rahachou}, finite element \cite{yamamoto}, and finite
difference time domain (FDTD) methods \cite{hagness}. In this work
we will use FDTD simulations which provide rigorous numerical
results but which are rather time-consuming and therefore not
adapt to scan large parameter ranges. We will thus resort to a
semi-analytical CMT as a fast tool for a detailed design parameter
analysis. These two methods complement one another and give a
powerful tool for the investigation of micro resonators. In
particular, we are interested in the system characteristics
dependent on disk diameter, waveguide width, gap width between the
linear waveguide and the disk resonator, and the surface quality
of the disk.


\section{Optical properties of waveguide-coupled whispering gallery modes}
\label{sec:wgm}

\subsection{Finite difference time domain calculations}

The first method we use to investigate the resonance behavior of
the microdisk is by FDTD simulations. Here, Maxwell's equations
are discretized in space and time, and the time evolution of the
electromagnetic fields is numerically calculated on this grid. To
obtain high accuracy it is necessary to make the cells much
smaller than the optical wavelength, which leads to long
calculation times, in particular for a three-dimensional (3D)
model.

In two dimensions it is possible to perform direct FDTD
simulations for relatively large disks of diameter $>50\mu$m. In
3D, simulations are only feasible for small disk diameters.
However, comparisons of 2D and 3D simulations for small disk
diameters have shown reasonable agreement. In the rest of this
paper, we will thus restrict ourselves to simplified 2D
calculations, where the disk and waveguide modes are calculated
for a geometry infinitely extended in the vertical ($y$)
direction. The 3D modes are assumed to be simple slices of
thickness $d_y$ of these infinite modes. For the calculations
presented in this work we always assume $d_y=5\mu$m. We will also
limit the calculations to electric fields polarized along the $y$
direction, i.e., to TE modes only.

In this case, Maxwell's equations for the electric and magnetic
fields, $E_y$, $H_x$, and $H_z$ reduce to
 \bea
 \frac{\partial{H}_{x}}{\partial{t}} &=&
    -\frac{1}{\mu}_{0}\frac{\partial{E}_{y}}{\partial{z}},
    \label{eq:maxwell1}\\
 \frac{\partial{H}_{z}}{\partial{t}} &=&
    -\frac{1}{\mu}_{0}\frac{\partial{E}_{y}}{\partial{x}},
    \label{eq:maxwell2}\\
 \frac{\partial{E}_{y}}{\partial{t}} &=&
    -\frac{1}{\varepsilon}(\frac{\partial{H}_{z}}
    {\partial{x}}-\frac{\partial{H}_{x}}{\partial{z}}),
    \label{eq:maxwell3}
 \eea
where $\mu_0$ and $\varepsilon$ are the magnetic and electric
permeability, respectively. The FDTD simulations solve Eqs.\
(\ref{eq:maxwell1})-(\ref{eq:maxwell3}) on a spatial grid for a
given initial field distribution. We used the so-called unsplit
perfectly matched layer (UPLM) boundary conditions \cite{berenger}
and a uniform discretization with a step size range of 0.02$\mu$m
to 0.08$\mu$m in the $x$ and $z$ directions and with a time step
range of $4.45\times{10}^{-17}$s to $1.78\times{10}^{-16}$s.

For our calculations the initial condition was such that the disk
was empty and that light was pumped into the lowest transverse
mode of the linear waveguide from one end. The incoming light was
either a continuous wave or a 30fs Gaussian pulse. In the former
case, we are interested in the steady state field distribution
which, for example, allows us to observe the resonant disk mode
and the evanescent field. Pulsed input allows the calculation of
the output power at the other end of the linear waveguide as a
function of frequency, i.e., the transmission spectrum, by
applying a discrete Fourier transform on the output field,
calculating the Poynting power density and integrating over the
waveguide cross-section.


\subsection{Coupled mode theory}
\label{sec:CMT}

We complement the numerical FDTD results with a semi-analytical
CMT. For this, it is assumed that the linear waveguide supports
only a single transverse mode, while the disk supports two
degenerate, counter-propagating WGM. The pumped waveguide mode
only couples to the forward propagating disk mode, but scattering
due to side-wall roughness may couple light into the second mode.
Both WGM are coupled to the waveguide mode via their evanescent
fields at point 1 in Fig.\ \ref{fig:structure}. For the
calculations we closely follow the work by Rowland and Love
\cite{love}.

First, the WGM are obtained as solutions of Eqs.\
(\ref{eq:maxwell1})-(\ref{eq:maxwell3}) in cylindrical
coordinates. This gives mode functions of the form
 \be
   E_{WGM}(r,\phi) = \left\{
   \begin{array}{ll}
   \frac{J_l(kn_cr)}{J_l(kn_cR)} e^{\pm il\phi}
   & \quad \mbox{ for } r<R \\
   \frac{H_l^{(1)}(kn_{cl}r)}{H_l^{(1)}(kn_{cl}R)} e^{\pm il\phi}
   & \quad \mbox{ for } r>R
   \end{array}
   \right.
 \ee
where $J_l$ are Bessel functions and $H_l^{(1)}$ are Hankel
functions of the first kind, and $R$ is the disk radius. The
eigenvalue equation for these modes is given by
 \be
 n_c\frac{{J}_{l+1}(kn_cR)}{{J}_{l}(kn_cR)}
 = n_{cl}\frac{H^{(1)}_{l+1}(kn_{cl}R)}{H^{(1)}_{l}(kn_{cl}R)}.
 \ee
As all WGM are lossy, the eigenvalues are complex
 \be
   k=k_r-ik_i
 \ee
and the intrinsic quality factor of the WGM is given by
\cite{love}
 \be
    Q_{WGM}=\frac{k_r}{2k_i}.
    \label{eq:Qwgm}
 \ee
Similarly, the mode functions $E_{lin}$ of the linear waveguide
are calculated for the same wavenumber $k$.

The second step of the CMT is to write the total light field as a
superposition of disk and waveguide mode
 \be
   E(x,z)=a_1(z){E}_{lin}(x,z) + a_2(z){E}_{WGM}(x,z).
 \ee
The coupled mode equations read
 \bea
  \frac{da_1}{dz} &=& -i{\beta}_{lin}a_1 +
  i{C}(z)a_2, \label{eq:cm1}\\
  \frac{da_2}{dz} &=&
  -i{\beta}_{WGM}(z)a_2+i{C}(z)a_1, \label{eq:cm2}
 \eea
where $\beta_{lin}$ and $\beta_{WGM}$ are the propagation
constants of the waveguide and the WGM, respectively, and $C(z)$
is the position-dependent coupling coefficient obtained by
calculating the mode overlap of waveguide and disk. For details of
this calculations see Ref.\ \cite{love}.

Finally, integrating Eqs.\ (\ref{eq:cm1}) and (\ref{eq:cm2}) over
$z$ in the region of significant coupling around point 1 in Fig.\
\ref{fig:structure} yields the coupler transmission matrix $T$
which relates the waveguide and disk output fields to the inputs,
 \be
  \left(\begin{array}{c}
  a_1\\a_2
  \end{array}\right)_{out} = T
  \left(\begin{array}{c}
  a_1\\a_2
  \end{array}\right)_{in}
 \ee
where
 \be
  T =
  \left(\begin{array}{cc}
  t_{11} & t_{12} \\ t_{21} & t_{22}
  \end{array}\right).
 \ee

The cavity decay rate (half width at half maximum) $\kappa_T$ of
the WGM due to the coupling to the waveguide is
 \be
 \kappa_T = |t_{12}|^2/(2T_r)
 \ee
where $T_r=2\pi l/(ck)$ is the round trip time. The corresponding
quality factor is
 \be
 Q_{coup} = ck/(2\kappa_T) = 2\pi l/|t_{12}|^2.
 \label{eq:Qcoup}
 \ee


\subsection{System losses}
\label{sec:losses}

Apart from the intrinsic WGM losses (\ref{eq:Qwgm}) and the
coupling losses into the waveguide (\ref{eq:Qcoup}) at least two
other loss mechanisms have to be taken into account, intrinsic
material losses and surface scattering losses.

The main material loss mechanisms are bulk Rayleigh scattering and
ultraviolet and infrared absorption. The corresponding quality
factor ${Q}_{mat}$ can be expressed in the form
\cite{gorodetsky,vernooy}
 \begin{equation}
     {Q}_{mat} = \frac{2n_c\pi}{\alpha\lambda},
 \end{equation}
where $\alpha$ is the loss coefficient and $\lambda=2\pi/k$ is the
vacuum wavelength. Material losses for fused silica in the
wavelength range near 780nm are of the order of 5dB/km, which
gives rise to $Q_{mat} \sim 10^{10}$.

Greater uncertainty is associated with losses due to surface
scattering and absorption due to surface roughness or the presence
of an absorbing impurity on the surface of the disk. For a given
size of surface inhomogeneities (roughness) and correlation
length, the surface scattering quality factor ${Q}_{surf}$ has to
take into account not only the direct scattering of light out of
the disk ("leakage") but also scattering into other modes with
high rates of leakage. Various expressions have been used to
describe quality limits due to surface scattering
\cite{gorodetsky,vernooy,banna}. In this work, we apply the
expression based on the model of Rayleigh scattering by
molecule-sized surface clusters \cite{gorodetsky}
 \begin{equation}
     {Q}_{surf} = \frac{D\lambda^{2}}{2L_{c}\pi^2\sigma^2},
 \end{equation}
where $D$ is the disk diameter, $\sigma$ is the root-mean-square
of the surface roughness and $L_{c}$ the surface correlation
length. As was reported in \cite{lee}, the numerical values for
$\sigma$ and $L_{c}$ can be less than 1nm and 5nm, respectively.
In our calculations we used $\sigma$=1nm, $L_{c}$=5nm and
$\sigma$=2nm, $L_{c}$=10nm.

The overall cavity quality factor taking all the loss mechanisms
discussed above into account is then given by
 \be
 Q^{-1} = Q_{coup}^{-1} + Q_{WGM}^{-1} + Q_{mat}^{-1} +
 Q_{surf}^{-1}.
 \label{eq:Qtotal}
 \ee
This is related to the cavity line width (HWHM) $\kappa$ by
 \be
 Q = ck/(2\kappa)
 \ee
and to the cavity finesse $F$ by
 \be
 F = Q\frac{\mathrm{FSR}}{ck} \approx Q/l
 \ee
where $\mathrm{FSR}\approx ck/l$ denotes the free spectral range
($l$ is the mode index of the WGM). For later use we also
introduce $\kappa_{loss}=\kappa-\kappa_T$ which is the cavity loss
rate due to losses into all channels apart from the linear
waveguide.


\subsection{Results}
\label{sec:results}

After describing the building blocks of our calculation, let us
now discuss some numerical results for the optical properties of
our system. This will serve two purposes: first, to compare our
different calculation methods with each other and with available
experimental data in order to show the accuracy of our results;
second, we need to apply our calculation to the experimental
parameters that are of interest in our case, in order to establish
a base for the atom-light interaction that will be discussed in
the following sections.

\begin{figure}
  \includegraphics[width=6.5cm]{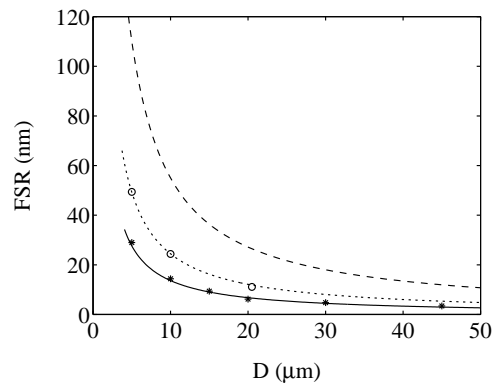}
  \caption{Free spectral range versus disk diameter. The lines present
  results of analytical calculations at wavelengths 780nm (solid line) and
  1550nm (dashed) in fused silica and at 1550nm for
  $n_c=3.2$ (dotted). Corresponding FDTD results are indicated by ($*$),
  and experimental data \cite{hagness} by ($\circ$).
  \label{fig:FSR}}
\end{figure}

We have compared the spatial profile of the WGM obtained from FDTD
simulation with the one resulting from a CMT calculation, and
found good agreement. We have also looked at the resonance
spectrum of the disk for different parameters, and extracted
values for the free spectral range (FSR) and the quality factor
$Q$. As presented below, here too the agreement was good.

In Fig.\ \ref{fig:FSR}, we plot the FSR for disk diameters $D$ in
the range of 5 - 50$\mu$m, refractive index values 1.454 - 3.2 and
wavelengths near 0.78$\mu$m and 1.55$\mu$m. The latter is chosen
to compare with previously published results \cite{hagness}. FDTD
simulations and analytically calculated WGM are in excellent
agreement with each other and with the experimental data. As
expected, the FSR is approximately inversely proportional to the
disk diameter.

\begin{figure}
\includegraphics[width=6.5cm]{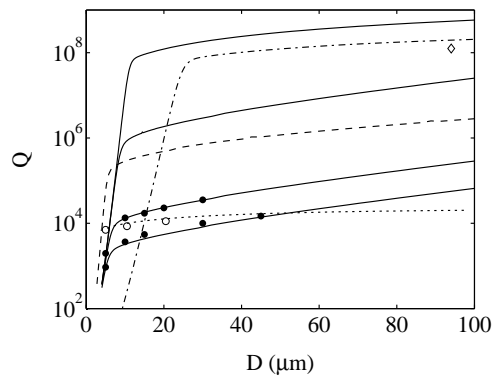}
\caption{Quality factor $Q$ versus disk diameter for various gap
sizes and materials. Solid curves (from bottom to top): CMT
results for gap sizes $0.1\mu$m, $0.2\mu$m, $0.5\mu$m, and for the
uncoupled disk for $\lambda=780$nm, $n_c=1.454$, $\sigma=1$nm,
$L_c=5$nm, ($\bullet$) represent FDTD simulations. Dashed curve:
$n_c=2.17$, gap size $0.2\mu$m. CMT results for $\lambda=1550$nm,
$\sigma=2$nm, $L_c=10$nm: uncoupled disk with $n_c=1.444$
(dash-dotted curve), $n_c=3.2$ and gap size $0.1\mu$m (dotted
curve). Corresponding experimental data is given by ($\diamond$)
\cite{Kippenberg} and ($\circ$) \cite{hagness}, respectively.
\label{fig:Qtotal}}
\end{figure}

Figure \ref{fig:Qtotal} shows the total cavity quality factor $Q$,
Eq.\ (\ref{eq:Qtotal}), dependent on the outer diameter $D$ of the
disk and the size of the air gap between disk and linear
waveguide. $Q$ was calculated using FDTD and CMT models, which
again show excellent agreement. We note that very high quality
factors up to about $10^8$ can be achieved with the current system
for disk diameters of several tens of microns. The upper limit for
$Q$ shown in Fig.\ \ref{fig:Qtotal} is the value obtained for a
very large air gap, where coupling losses $\kappa_T$ are
negligible and the cavity quality is limited by the intrinsic and
material quality factors $Q_{WGM}$, $Q_{mat}$, and $Q_{surf}$.

Figure \ref{fig:Qtotal} also shows data for two different
wavelengths and varying refractive index. Note that $Q$ generally
increases with increasing refractive index due to the better
confinement and therefore weaker coupling to the waveguide.
Similarly, $Q$ decreases with increasing wavelength due to
stronger waveguide coupling. Nevertheless, for each case there are
unique material losses and fabrication uncertainties which may
change this general behavior.

To calculate the $Q$ factor for different disk sizes, the index
$l$ of the WGM must be changed accordingly to keep the resonant
wavelength near 780nm relevant for Rb atoms. The necessary
wavelength for optimal mode resonance can be achieved by choosing
a precise disk size. We will discuss tuning of the micro resonator
later in Sec.\ \ref{sec:experiment}. Note also that, while the
intrinsic quality factor $Q_{WGM}$ is always lower for higher
radial modes, the overall $Q$ can in fact be higher due to weaker
coupling (smaller losses) to the linear waveguide, which is the
dominating loss mechanism for most parameter regimes we are
interested in. Table \ref{table1} shows the optical properties of
a few selected disk-waveguide geometries.

\begin{table}
 \begin{ruledtabular}
 \begin{tabular}{ccccccc}
 $D$ ($\mu$m) & $l$ & $q$ & $\lambda$ (nm) & $Q_1$ & $Q_2$ & $g_0$ (MHz)\\
  30 & 167 & 1 & 778.73 & $1.55\times 10^5$ & $8.44\times 10^6$ & 102.6  \\
  30 & 166 & 1 & 783.27 & $1.47\times 10^5$ & $8.05\times 10^6$ & 103.2 \\
  30 & 159 & 2 & 780.04 & $1.83\times 10^5$ & $8.85\times 10^6$ & 102.8 \\
  15 & 81 & 1 & 780.41 & $7.66\times 10^4$ & $3.82\times 10^6$ & 205.7\\
  45 & 253 & 1 & 780.15 & $2.66\times 10^5$ & $1.40\times 10^7$ & 68.5\\
 \end{tabular}
 \end{ruledtabular}
\caption{Optical properties of selected WGM. $Q_1$ ($Q_2$) is the
quality factor $Q$ for a gap size of 0.3$\mu$m (0.6$\mu$m), $l$
and $q$ are the longitudinal and radial mode index, respectively,
$g_0$ is the single-photon Rabi frequency for an atom at the disk
boundary. Surface parameters are $\sigma=2$nm, $L_c=10$nm. Results
are obtained using CMT. \label{table1}}
\end{table}


\section{The Jaynes-Cummings model}
\label{sec:coupling}

In this section we introduce a model for the coupling of a
two-level atom to the light field outside of the microdisk. We
consider two coherent modes with complex amplitudes $\alpha_+$ and
$\alpha_-$ traveling in opposing directions, and assume atom-light
coupling to these modes via the respective single-photon Rabi
frequencies $g_+$ and $g_-$. The coupling of the two modes due to
imperfections in the disk is accounted for by the introduction of
the complex coefficient $\epsilon$, and the model includes the
possibility of pumping from either direction with rates $\eta_+$
and $\eta_-$, where $\eta_\pm=t_{21}A_{in,\pm}/\sqrt{T_r}$. Here
$A_{in,\pm}$ is the amplitude of the pump field in each direction
within the waveguide such that $|A_{in,\pm}|^2$ is the photon flux
in units of photons per second, and $t_{21}$ and $T_r$ have been
defined in Sec.\ \ref{sec:CMT}.

Ignoring the external motion of the atom, the Hamiltonian of this
system can be written as $(\hbar=1)$
 \bea
 \nonumber H& =& -\Delta_a \sigma_{11} - \Delta_c(a_+^\dagger a_- +a_-^\dagger a_-)\\
 \nonumber  &  & -i(g_+ a_+^\dagger\sigma_{01}-g_+^*\sigma_{10}a_+)- i\eta_+(a_+ - a_+^\dagger)\\
 \nonumber  &  & -i(g_- a_-^\dagger\sigma_{01}-g_-^*\sigma_{10}a_-)- i\eta_-(a_- - a_-^\dagger)\\
            &  & -i(\epsilon a_+^\dagger a_- - \epsilon^* a_-^\dagger a_+)\:,
 \eea
where $\Delta_a$ and $\Delta_c$ are the atomic and resonator
detunings, respectively, $\sigma_{10}$ and $\sigma_{01}$ the
atomic raising and lowering operators, and $a_\pm^\dagger$ and
$a_\pm$ the mode creation and annihilation operators. Here the
energy of the lower atomic state has been set to zero. The
equation of motion for the density operator of the system can be
written as
 \be
 \frac{d}{dt}\rho=-i[H,\rho]+L\rho
 \ee
where $L$ is the usual linear operator describing cavity and
atomic decay with rates $\kappa$ and $\Gamma$ respectively.

Assuming a factorized density operator $\rho$, we find the
equations of motion for the elements of the atomic density
operator
 \bea
 \nonumber \frac{d}{dt}\rho_{10} & = & (-\Gamma+i\Delta_a)\rho_{10}\\
    && + (g_+^*\alpha_+ +g_-^*\alpha_-)(\rho_{00}-\rho_{11}),\label{eq:rho10}\\
 \nonumber \frac{d}{dt}\rho_{11} & = & -2\Gamma\rho_{11} + (g_+^*\alpha_+ + g_-^*\alpha_-)\rho_{01}\\
    && + (g_+\alpha_+^* + g_-\alpha_-^*)\rho_{10},\label{eq:rho11}
 \eea
and the coherent state amplitudes $\alpha_\pm$ obey the equations
of motion
 \bea
 \frac{d}{dt}\alpha_+& =& (i\Delta_c-\kappa)\alpha_+ - g_+ \rho_{10}
    + \eta_+ -\epsilon\alpha_- ,\label{eq:alp} \\
 \frac{d}{dt}\alpha_-& =& (i\Delta_c-\kappa)\alpha_- - g_- \rho_{10}
    + \eta_- +\epsilon^*\alpha_+.\label{eq:alm}
 \eea

In this work we are only interested in the stationary solution of
these equations of motion. To this end, we first solve the linear
set of equations (\ref{eq:alp}) and (\ref{eq:alm}) with respect to
$\alpha_\pm$. The resulting expressions for $\alpha_\pm$ are
linear in $\rho_{10}$ and can be inserted into (\ref{eq:rho10}).
From this and using $\rho_{11}+\rho_{00}=1$, we obtain $\rho_{10}$
as a function of $\rho_{11}$. That and the corresponding results
for $\alpha_\pm$ can be inserted into (\ref{eq:rho11}) to give a
real-valued nonlinear equation in $\rho_{11}$ which can be solved
by standard numerical techniques. The output field of the linear
waveguide is the superposition of the input field transmitted
through the waveguide-disk coupler and the light coupled out of
the disk,
 \be
 A_{out,\pm} =
 t_{11}A_{in,\pm}+\frac{t_{12}}{\sqrt{T_r}}\alpha_\pm.
 \ee

\begin{figure}
\includegraphics[width=6.5cm]{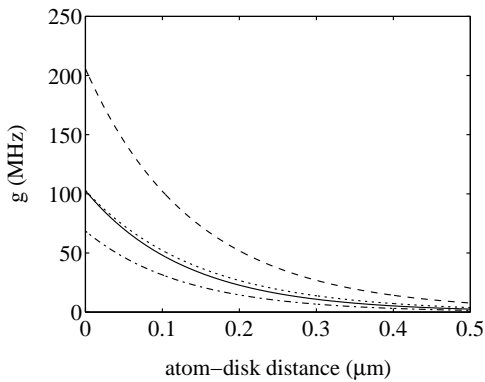}
\caption{Rabi frequency versus atom-disk distance. Solid curve:
$D=30\mu$m, mode indices $(l,q)=(167,1)$; dotted: $D=30\mu$m,
$(l,q)=(159,2)$; dashed: $D=15\mu$m, $(l,q)=(81,1)$; dash-dotted:
$D=45\mu$m, $(l,q)=(253,1)$. \label{fig:Rabi}}
\end{figure}

The Rabi frequency $g_\pm$ is given by
 \be
 g_\pm = E_{WGM}(\mathbf{x}_{a})\left[
   \frac{3\Gamma c^3}{\omega^2 d_y \int r n(r)^2 |E_{WGM}(r)|^2 dr}
   \right]^{1/2}
 \ee
where $n(r)=n_{c}$ ($n_{cl}$) for $r<R$ ($r>R$), $d_y$ is the disk
height, $\omega$ is the atomic transition frequency, and
$\mathbf{x}_a=(r_a,\phi_a)$ is the atomic position in the
evanescent field of the disk modes. Figure \ref{fig:Rabi} shows
the Rabi frequency depending on the distance of the atom from the
disk surface for several disk diameters and mode numbers. The
maximum values for $g_\pm$ at the disk surface for several
selected modes are also given in Table \ref{table1}.

We note that for efficient atom-mode coupling the atom should be
placed within 50-100nm from the disk. Such close proximity should
be feasible without the atom being absorbed by the surface, as the
van der Waals forces may be counter balanced by blue detuning of
the resonator light \cite{kimble}. In addition, such position
resolution is allowed by the 10nm and better ground state sizes
achievable in magnetic traps on the atom chip \cite{folman}.
Future work will include exact simulations of the near surface
potential.


\section{Single-atom detection}
\label{sec:detection}

In the following we will investigate the properties of a
microdisk, modeled as a high finesse ring resonator as described
in the previous section, as a single-atom detector for quantum
information processing on an atom chip. The scheme we discuss here
is based on homodyne detection of the phase change of the light at
the output of the linear waveguide in the forward direction. There
are several advantages of this scheme over a corresponding
absorption detection. (i) It allows one to drive the atom far off
resonance, in which case the precise tuning of the disk resonator
with respect to the atomic transition frequency is of minor
importance. Cavity tuning will be discussed in more detail in
Sec.\ \ref{sec:experiment}. (ii) If the additional loss mechanisms
discussed in Sec.\ \ref{sec:losses} are small compared to the
disk-waveguide coupling strength, all of the pump light will leave
the system through the forward waveguide output. Therefore for
most parameter regimes a strong signal can be expected, which
allows the use of standard photodetectors rather than
sophisticated single-photon counters. (iii) The strong output
signal also provides stability of the detection scheme against
weak background scattering processes.

Our detection scheme works as follows. The output field of the
straight waveguide is mixed with a strong local oscillator field
at a 50-50 beamsplitter and the light intensities in the two
beamsplitter outputs are measured and integrated over the
observation time $\tau$ to give the total number of detected
photons $N_1$ and $N_2$. The signal we are interested in is given
by the difference $|N_1-N_2|$. The phase of the local oscillator
is adjusted such that this difference is zero when no atom is
interacting with the disk field. The presence of an atom is then
inferred from a change in this intensity difference. Assuming that
the detection is shot-noise limited, the signal-to-noise ratio $S$
of the atom detection is given by
 \be
 S = \frac{|N_1-N_2|}{\sqrt{N_1+N_2}}
 \approx 2\sqrt{\tau}|A_{out,0}||\sin(\phi-\phi_0)|
 \label{eq:S}
 \ee
where $\phi$ ($\phi_0$) is the phase of the waveguide output field
$A_{out,+}$ with (without) an atom, $A_{out,0}$ is the field
amplitude without an atom, and we have assumed that only the phase
and not the amplitude of the output is changed due to the atom.

The second quantity of interest is the number of photons $M$
spontaneously scattered by the atom during the interaction time,
given by
 \be
 M = 2\Gamma\tau\rho_{11}.
 \label{eq:M}
 \ee
This should be as small as possible in order to minimize the
backaction of the detection process onto the atom. Ideally, $M\ll
1$ corresponds to the limit of non-destructive measurements.
Finally, we note that $S$ and $M$ scale differently with
interaction time $\tau$. For any given parameters, we can thus
choose $\tau$ in such a way to yield a certain, fixed
signal-to-noise ratio. As an example, we will in the following
consider the number $M_{10}$ of spontaneously scattered photons
when $\tau$ is rescaled to give $S=10$,
 \be
 M_{10} = 100 M/S^2.
 \label{eq:M10}
 \ee

For simplicity we will assume that the cavity is driven on
resonance with the disk modes, $\Delta_c=0$, thereby minimizing
the effect of other, off-resonant modes. The atom is assumed to be
far off resonance with respect to the cavity mode,
$\Delta_a\gg\Gamma$, such that the effect of the atom is mainly to
provide a phase shift of the cavity mode. Under these conditions
and in the limit of small atomic saturation we can derive
analytical approximations for $S$, $M$ and $M_{10}$ \cite{horak}
 \bea
 S &=& 4\sqrt{\tau}|A_{in}|\frac{\kappa_T g^2}{\Delta_a\kappa^2},
 \label{eq:sapprox}\\
 M &=& 4\tau|A_{in}|^2\frac{\kappa_T g^2\Gamma}{\Delta_a^2\kappa^2},
 \label{eq:mapprox}\\
 M_{10} &=& 25 \frac{\kappa^2\Gamma}{\kappa_T g^2}.
 \label{eq:m10approx}
 \eea
Note that $M_{10}$ is independent of the pump power and of the
atomic detuning.

\begin{figure}
  \includegraphics[width=6.5cm]{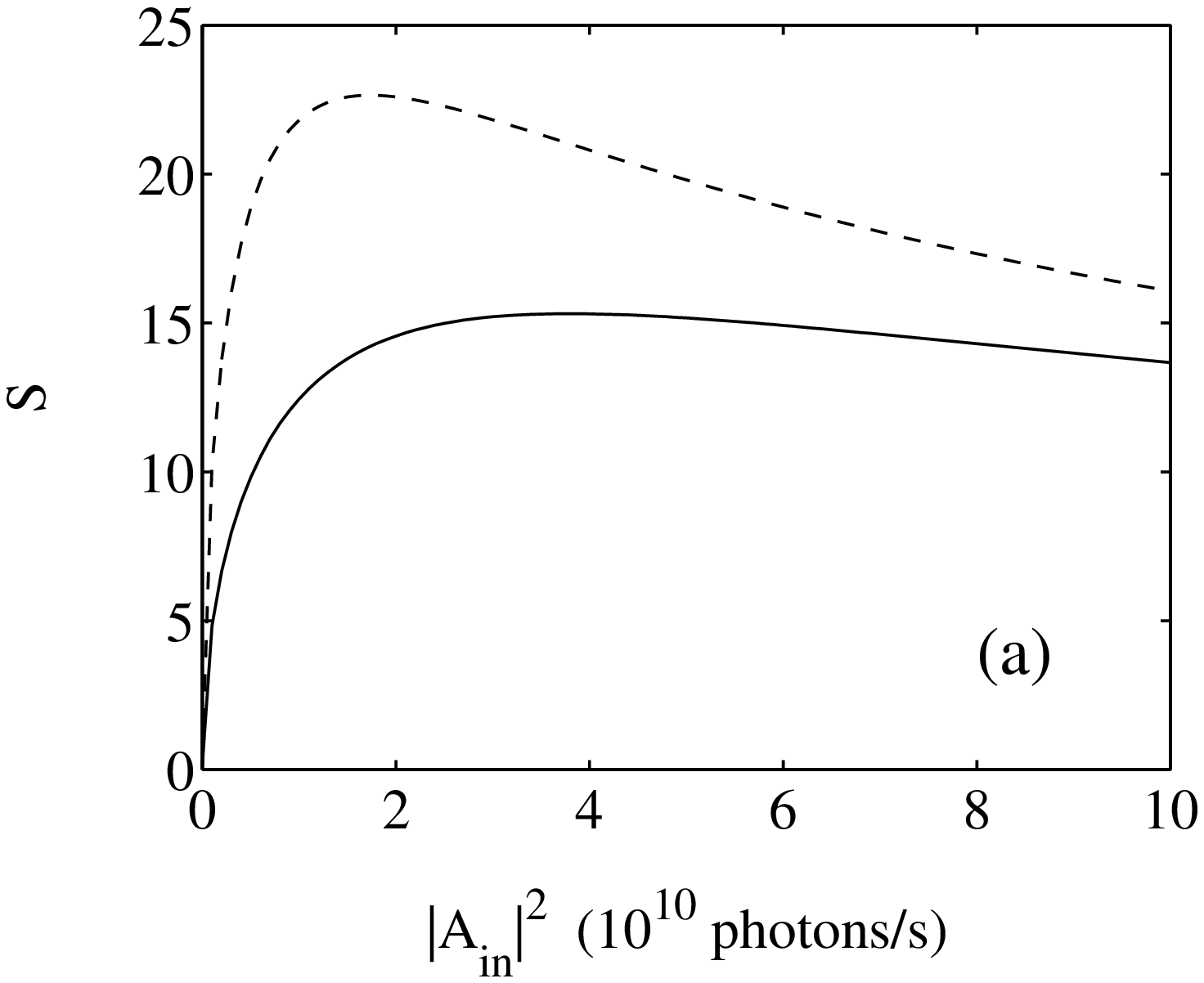}
  \includegraphics[width=6.5cm]{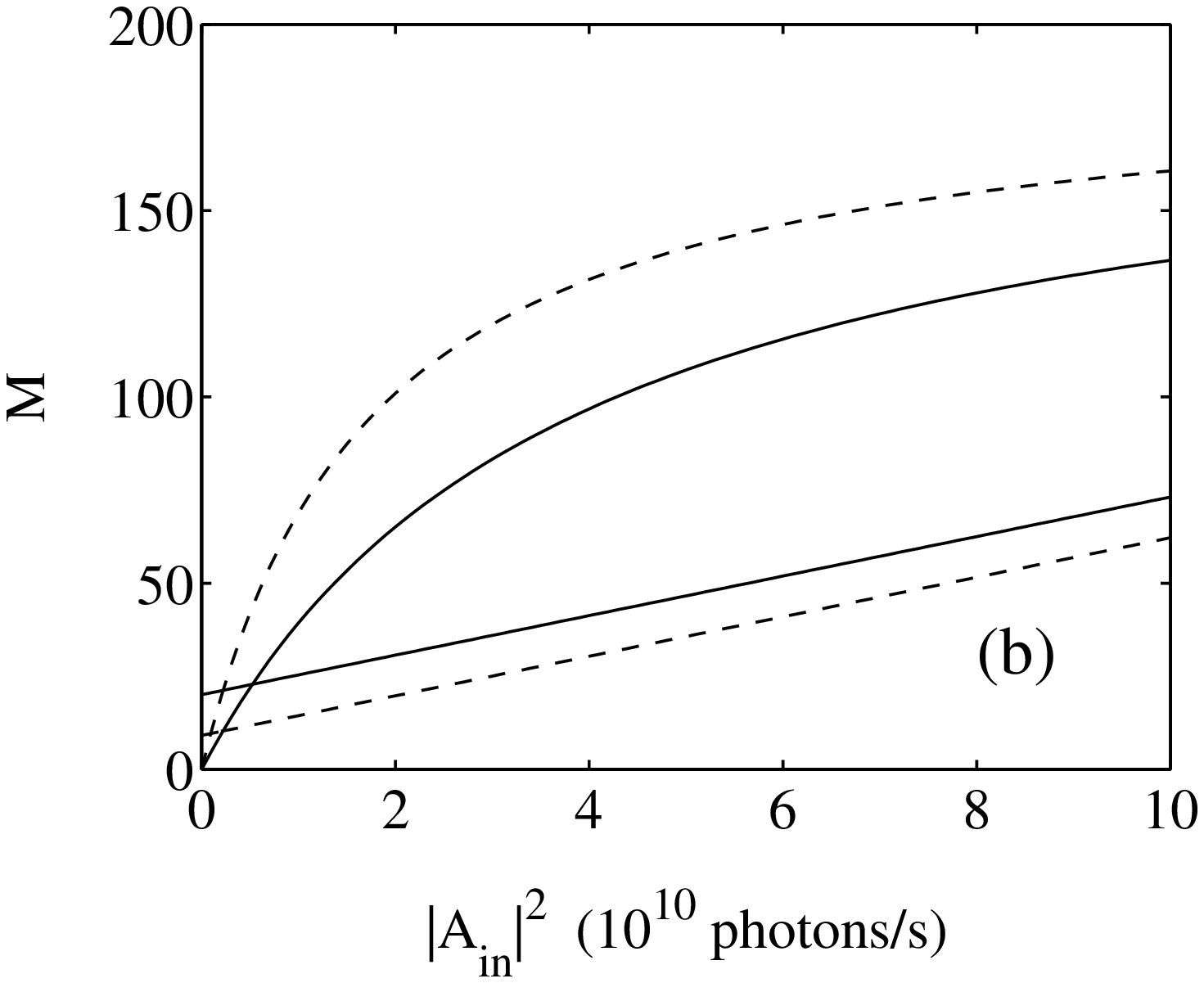}
  \caption{(a) Signal-to-noise ratio versus pump intensity for disk
  diameter 30$\mu$m (solid line) and 15$\mu$m (dashed). (b)
  Corresponding photon scattering $M$ (top curves) and $M_{10}$
  (bottom). Gap size is 0.3$\mu$m, waveguide width is 0.6$\mu$m,
  $\sigma=2$nm, $L_c=10$nm, $\Delta_a=100\Gamma$, $\tau=10\mu$s,
  and the atom is assumed to sit 50nm away from the disk surface.
  \label{fig:s_vs_pump}}
\end{figure}

In Fig.\ \ref{fig:s_vs_pump} we show $S$, $M$ and $M_{10}$ as a
function of the input power to the linear waveguide. For a weak
pump the signal-to-noise ratio increases with power since more
photons are coupled into the cavity and interact with the atom.
However, because of saturation the atom can only interact with a
maximum number of photons in a given interaction time. Hence, $S$
reaches a maximum value, and for stronger pump powers $S$ is
decreasing again. The number of spontaneously scattered photons
$M$ increases with pump power and finally saturates at the value
$\Gamma\tau$. $M_{10}$ shows an approximately linear increase with
pump power, indicating that the least perturbing atom detection
for a given value of $S$ is achieved for low atomic saturation.
There is, however, a trade-off as the required interaction time
increases in this limit. For low saturation, $\rho_{11}\ll 1$, the
numerical results are accurately described by the approximations
(\ref{eq:sapprox})-(\ref{eq:m10approx}).

\begin{figure}
  \includegraphics[width=6.5cm]{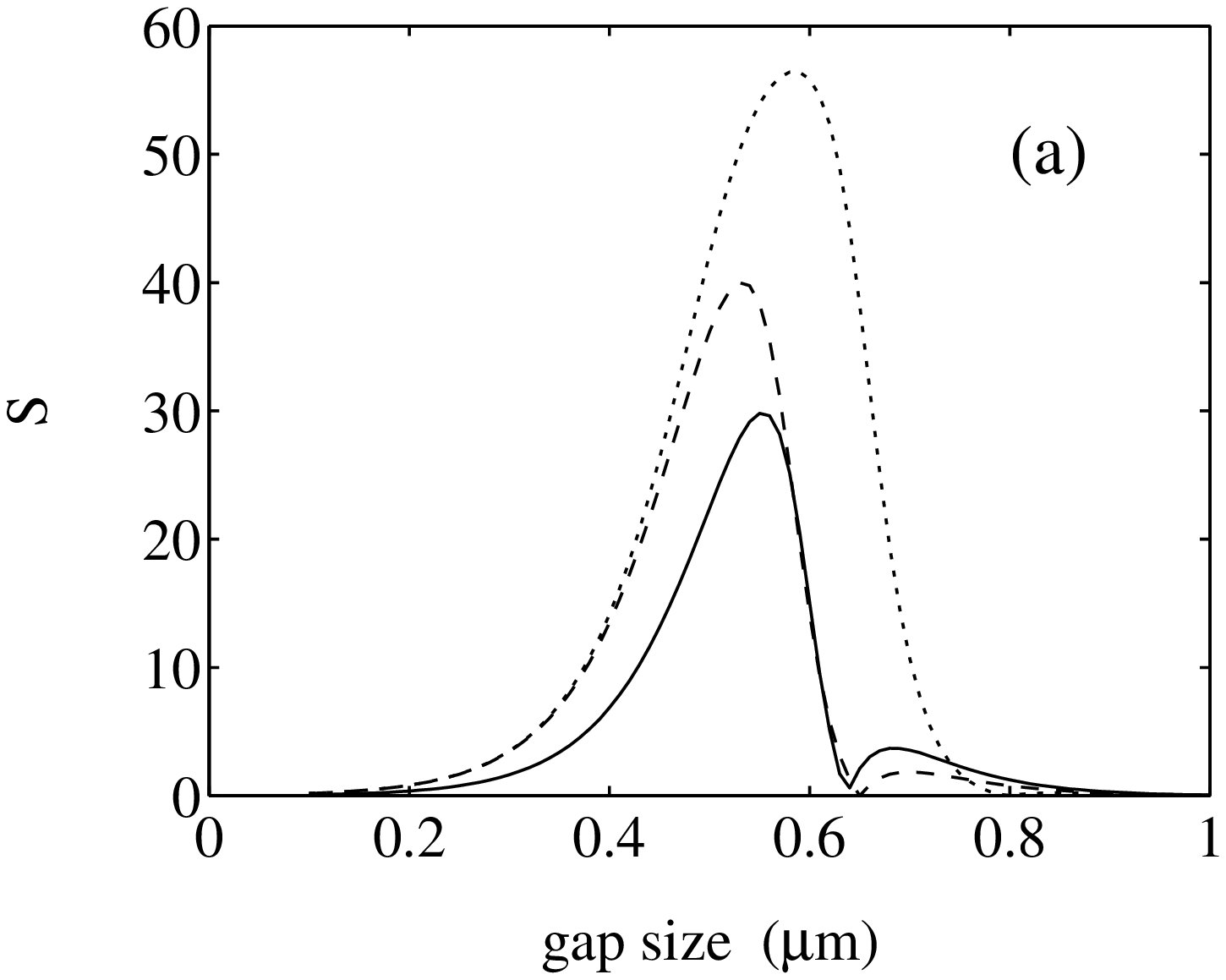}
  \includegraphics[width=6.5cm]{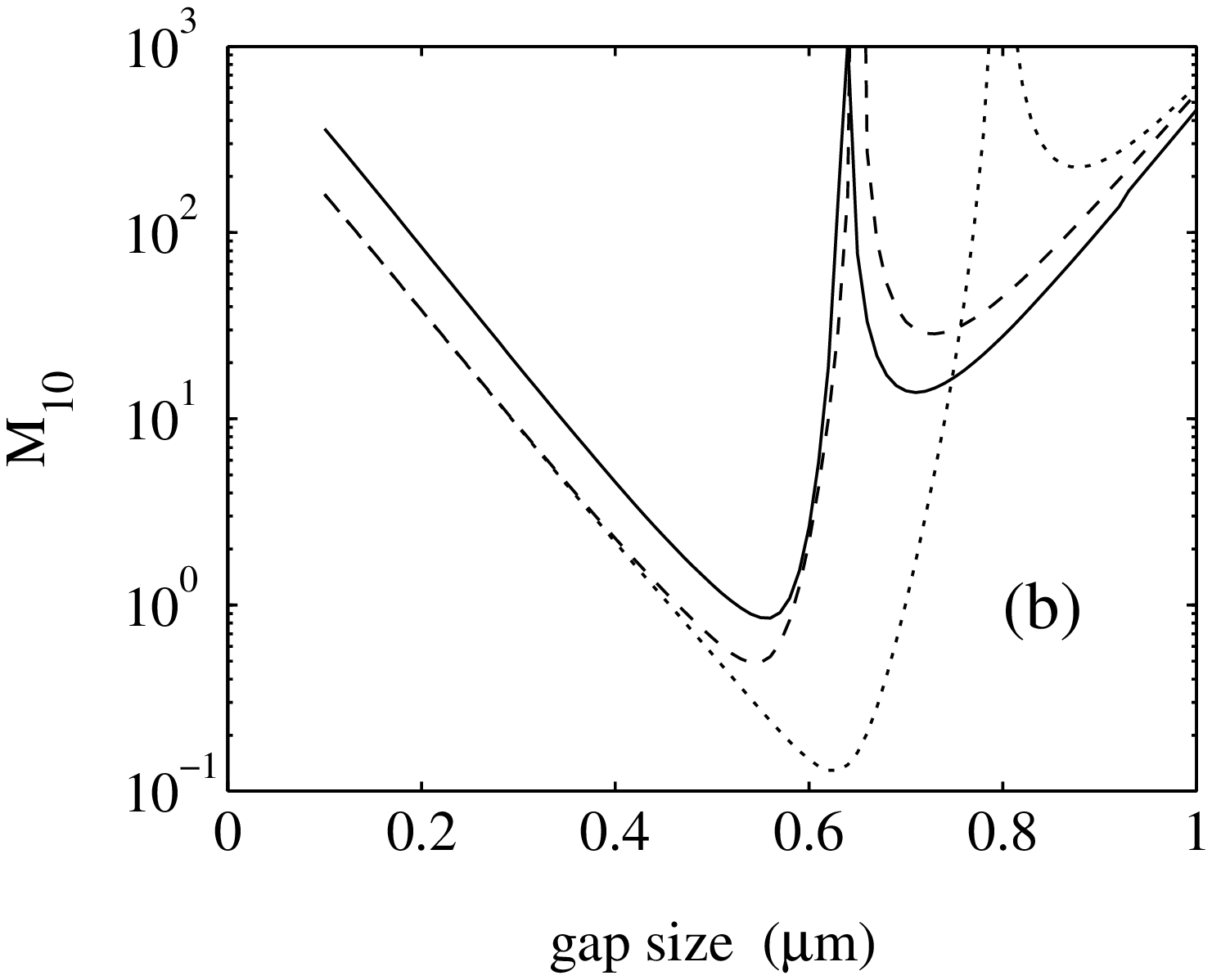}
\caption{(a) Signal-to-noise ratio $S$ and (b) scattered photons
$M_{10}$ versus gap size for weak pumping ($|A_{in}|^2=10^8$
photons/s). Solid curve: $D=30\mu$m, dashed: $D=15\mu$m for
$\sigma=2$nm, $L_c=10$nm. Dotted curve: $D=15\mu$m, $\sigma=1$nm,
$L_c=5$nm. Waveguide width is 0.6$\mu$m, distance atom-disk is
50nm, interaction time is $10\mu$s. \label{fig:s_vs_gap}}
\end{figure}

Figure \ref{fig:s_vs_gap} shows $S$ and $M_{10}$ as a function of
the gap between the microdisk and the waveguide in the limit of
weak atomic saturation ($\rho_{11}<0.03$ for the shown parameter
range). For small enough gap sizes, increasing the gap reduces the
coupling between disk and waveguide modes and therefore the cavity
finesse increases. This leads to improved signal-to-noise ratios
and less spontaneous photon scattering by the atom. For very large
gaps, on the other hand, the cavity finesse is limited by the
additional losses, see Sec.\ \ref{sec:losses}. In this case,
increasing the gap even further reduces the number of photons
coupled back from the cavity into the waveguide and thus reduces
the detected signal. If the cavity-waveguide coupling exactly
equals the additional losses, no light is transmitted through the
waveguide at all, which leads to the points of $S=0$ and the
corresponding divergence of $M_{10}$ observed in the figure. We
find numerically that $S$ is maximum and $M_{10}$ is minimum if
losses from the disk into the waveguide are about 4-5 times higher
than the additional losses.

For the parameters of Fig.\ \ref{fig:s_vs_gap} we find that the
minimum value of $M_{10}$ is 0.85 for a disk diameter of
$D=30\mu$m (solid line) and 0.49 for $D=15\mu$m (dashed line). The
reason for this difference is mainly that for the smaller disk
more energy of the resonant mode is in the evanescent field. This
leads to improved coupling of the atom to the mode, that is, to a
larger Rabi frequency $g$ as already seen in Fig.\ \ref{fig:Rabi}.
For both disk diameters, however, the minimum value of $M_{10}$ is
below unity, which indicates that single atoms can be detected
while on average scattering less than one photon spontaneously.
Moreover, we observe that this minimum value of $M_{10}$ is
limited by the additional losses due to surface roughness. If the
surface parameters are decreased by a factor of two to
$\sigma=1$nm and $L_c=5$nm, $M_{10}$ can be as small as 0.13,
shown by the dotted curve in Fig.\ \ref{fig:s_vs_gap}(b). In this
case, our atom detection scheme approaches the limit of a
non-destructive measurement.

Note also that the parameters where optimum single-atom detection
is observed correspond to the strong coupling regime of cavity QED
\cite{QED}, defined by $g^2/(\kappa\Gamma)\gg 1$. In particular,
at the minimum points of $M_{10}$ in Fig.\ \ref{fig:s_vs_gap} we
find $g^2/(\kappa\Gamma)=48$ ($D=30\mu$m), 73 ($D=15\mu$m,
$\sigma=2$nm, $L_c=10$nm), and 290 ($D=15\mu$m, $\sigma=1$nm,
$L_c=5$nm).

\begin{figure}
\includegraphics[width=6.5cm]{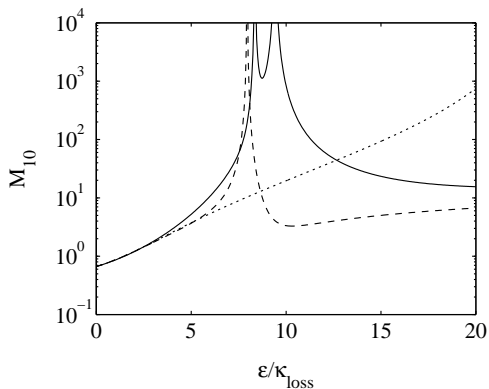}
\caption{$M_{10}$ versus mode coupling parameter
$\epsilon/\kappa_{loss}$ for $D=15\mu$m, gap size $0.5\mu$m,
$\sigma=2$nm, $L_c=10$nm, $\Delta_a=100\Gamma$, $|A_{in}|^2=10^8$
photons/s. Solid curve: $\Delta_c=0$, dashed: $\Delta_c=\epsilon$,
dotted: $\Delta_c=-\epsilon$. \label{fig:s_vs_epsil}}
\end{figure}

Finally, in Fig.\ \ref{fig:s_vs_epsil} we show the effect of mode
coupling between the two counter-propagating WGM on the
single-atom detection. For $\epsilon$ smaller than the total
cavity linewidth $\kappa$, the main effect of mode coupling is to
increase the loss rate of the forward propagating mode. This
decreases the mode quality factor and therefore increases
$M_{10}$. For $\epsilon>\kappa$, the strong coupling lifts the
degeneracy between the counter-propagating modes. The resulting
new frequency eigenmodes of the disk are shifted in frequency by
$\pm\epsilon$ with respect to the uncoupled modes. Tuning one of
these coupled modes into resonance with the pump may thus seem
advantageous. However, as can be seen from the corresponding
curves in Fig.\ \ref{fig:s_vs_epsil}, this is not the case for
some parameter regions. Several effects play a role here: (i) our
detection scheme only measures the forward propagating output of
the linear waveguide and thus is sensitive to the phase difference
between the two coupled modes, which in turn changes with the
frequency shift; (ii) coupling to the backward propagating mode
can be regarded as an additional loss, which shifts the position
of the divergences of $M_{10}$ already observed in Fig.\
\ref{fig:s_vs_gap}; (iii) strong coupling between the
counter-propagating modes fixes their relative phase, which
creates a standing-wave pattern at the disk surface. Thus, the
atom detection becomes dependent on the position of the atom. In
particular, if the atom is trapped at a node of the standing wave,
no interaction with the light occurs and atom detection is
impossible. The interplay between these effects explains the
features of Fig.\ \ref{fig:s_vs_epsil}. Note finally that for high
quality disk resonators only weak mode coupling is expected, for
example, in Ref.\ \cite{Kippenberg} a mode coupling parameter
$\epsilon/\kappa_{loss}$ of 1.5 has been reported.


\section{Experimental considerations and tunability}
\label{sec:experiment}

In this section we consider the feasibility of a microdisk
resonator in a realistic experiment on an atom chip. In particular
we discuss the effects of finite fabrication tolerances, of
temperature fluctuations, and various possibilities to tune the
resonance frequencies of the disk.

Let us first consider fabrication imperfections due to edge
roughness, which changes the disk diameter, or due to impurities,
which change the index of refraction. To first order, these
uncertainties affect the mode resonance through the simple
relation \cite{klitzing}
 \be
 \Delta\nu/\nu = -\Delta n/n - \Delta D/D,
 \label{eq:tune}
 \ee
where $\Delta\nu$, $\Delta n$, and $\Delta D$ are the changes of
the resonance frequency, the refractive index, and the disk
diameter, respectively. Assuming typical fabrication tolerances
for the disk diameter of a few nm and for the refractive index of
the order of $10^{-5}$ therefore leads to frequency shifts of the
order of tens of GHz.

Similarly, temperature fluctuations typically give rise to changes
of the refractive index of the order of $10^{-6}$ per $1^0$C. This
again will shift resonance frequencies of the microdisk by amounts
in the GHz regime. Therefore, methods to dynamically tune the disk
resonator have to be implemented on the chip in order to provide
frequency stability.

Tunable photonics is also needed if we want to trap, manipulate,
and measure different kinds of atoms or atoms in various internal
states. Accurate control of the light-matter interaction is
especially critical if one is to enable the exploitation of
quantum technology, for example, in the case of quantum
communication in order to convert a flying quantum bit (qubit) in
the form of a photon into a storage qubit (atom), in the case of
quantum computing where single qubit rotations, e.g., between two
hyperfine states, are needed, or in the case of sensors to measure
interference patterns.

As already noted, high $Q$ devices such as microspheres,
Fabry-Perot cavities, or microdisks are an extremely effective
tool for the delicate manipulation and measurement of subtle
quantum states, where a single photon can interact many times with
the same atom so that significant interaction can be achieved.
However, such strong coupling requires that the device is kept on
resonance with the exact frequency close to the chosen atom
transition \cite{horak,klitzing}.

In general, the FSR of a microdisk is orders of magnitude larger
than the linewidth, and therefore coincidences between the
fundamental transverse WGM and specific atomic transitions are
extremely unlikely. To keep a WGM resonance near the wavelength of
interest we thus need to change the diameter or the refractive
index of the disk. In order to achieve any given frequency, the
tuning range has to be of the order of the FSR. Furthermore, the
procedure should be stable, reversible, and fast enough to
compensate for temporal instabilities such as those arising from
temperature fluctuations of the chip. Under these conditions a
resonant mode close to the required atomic frequency can always be
found, even if the disk was initially fabricated with some
mismatch in diameter or in refractive index.

\begin{figure}
  \includegraphics[width=7cm]{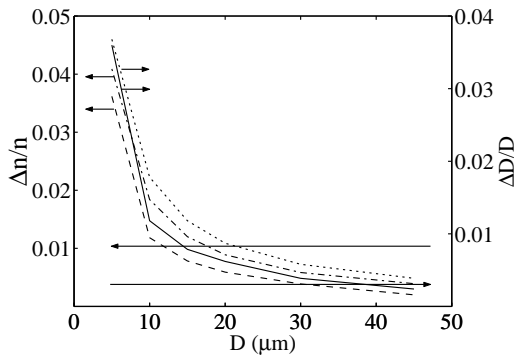}
  \caption{Required change of disk diameter and refractive index
  for a full FSR scan versus disk diameter at $\lambda=780$nm.
  $\Delta D/D$ and $\Delta n/n$ are presented by dotted and
  dash-dotted curves for $n=1.454$, and by solid and dashed curves
  for $n=2.17$, respectively. Typical values for actual materials
  are also presented by the long base arrows.
  \label{fig:Tdisk}}
\end{figure}

In order to scan a full FSR we require that $\Delta\nu/\nu \approx
1/l$, where $l$ is the longitudinal mode index. Together with Eq.\
(\ref{eq:tune}) this yields the required relative change of disk
diameter and refractive index. Results for this relative change
are shown in Fig.\ \ref{fig:Tdisk} versus the disk diameter for
fused silica ($n=1.454$ at $\lambda=780$nm) and for $n=2.17$ at
$\lambda=780$nm. Note however that since a few higher order radial
modes can also be used for some applications, there may in fact
exist several usable resonances within each FSR.

At least three possibilities may be considered to realize a
tunable microdisk resonator: UV, electro-optical, and piezo
tuning.

UV tuning was reported in \cite{javris}. There, it has been shown
that it is possible to change the refractive index of Ge doped
silica by up to $\Delta n=0.006$ using UV radiation. The drawback
is that this procedure is not reversible.

In the case of electro-optical tuning, the effect is achieved by a
uniform electric field that changes the refractive index and thus
allows one to tune the resonance wavelength. The idea is to cover
both the bottom and the top of the disk with a metal layer, and
apply a voltage to create the necessary electric field. For
example, electro-optic crystaline materials usually have a
relative change of refractive index of $0.01\%-1\%$ for a field of
$10^6$V/m (5V for our $5\mu$m disk thickness). Though demanding a
more complex fabrication process, such crystals may indeed be
used. Figure \ref{fig:Tdisk} shows that disk diameters as small as
$15\mu$m will enable a full FSR scan assuming $\Delta n/n=1\%$.
Other methods for such changes of the refractive index also exist.
For example, it was reported in \cite{djordjev} that tuning an InP
microdisk can be achieved utilizing free carrier injection, which
results in $\Delta n \sim 0.002$ up to $\Delta n\sim 0.01$.

By using a piezo effect one can change the diameter of the disk.
In this case, one has to fabricate the disk from a transparent
piezo material. The voltage necessary for tuning will be obtained
as above by two electrodes evaporated below and above the device.
Transparent piezo materials (such as BaTiO${}_3$) have a piezo
electric coefficient of order $10^{-10}$m/V which leads to $\Delta
D/D=0.003$ for electric fields of $30\times 10^6$V/m (150V for our
$5\mu$m disk thickness). Figure \ref{fig:Tdisk} shows that in this
case disk diameters as small as $30\mu m$ will enable a full FSR
scan. For both disk size tuning and refractive index tuning,
higher voltages and more exotic materials should allow for even
smaller disk sizes.


\section{Summary and conclusions}
\label{sec:conclusions}

We have investigated the feasibility of initiating strong
interaction between photons and single atoms on an atom chip by
utilizing microdisks as high $Q$ resonators.

Our calculations show that whispering gallery modes with high
quality factors of up to $10^8$ can be supported by disks with
diameters of several tens of microns. Single-atom detection with
high signal-to-noise ratios and practically no spontaneous photon
scattering can be achieved in such a system, and hence
non-destructive measurements may be possible.

We have also discussed different methods which would allow one to
tune the resonance frequencies of a microdisk over a full free
spectral range.

These results suggest that the regime of strong coupling between
atoms and photons can be achieved in optical resonators as small
as five microns in diameter. This could open the way for compact
arrays for quantum information processing.



\begin{acknowledgments}

This work was supported by the UK Engineering and Physical
Sciences Research Council and the European Union, Contract No.\
IST-2001-38863 (ACQP). S.H.\ is supported by a scholarship from
the Jersey Education Department. R.F.\ gratefully acknowledges the
support of the European Union FP6 'atomchip' collaboration, the
German Federal Ministry of Education and Research (BMBF) through
the DIP project, and the Israeli Science Foundation.

\end{acknowledgments}



\begin{thebibliography}{99}

\bibitem{slusher} R.\ E.\ Slusher, A.\ F.\ J.\ Levi, U.\ Mohideen,
S.\ L.\ McCall, S.\ J.\ Paerton, and R.\ A.\ Logan, Appl.\ Phys.\
Lett.\ \textbf{63}, 1310 (1993). D.\ Sadot and E.\ Boimovich, IEEE
Comm.\ Magazine \textbf{36} (12), 50 (1998).

\bibitem{vollmer} F.\ Vollmer, D.\ Braun, A.\ Libchaber, M.\
Khoshsima, I.\ Teraoka, and S.\ Arnold, Appl.\ Phys.\ Lett.\
\textbf{80}, 4057 (2002).

\bibitem{QED} T.\ Pellizzari, S.\ A.\ Gardiner, J.\ I.\ Cirac, and P.\ Zoller,
Phys.\ Rev.\ Lett.\ \textbf{75}, 3788 (1995); S.\ J.\ van Enk, J.\
I.\ Cirac, and P.\ Zoller, Science \textbf{279}, 205 (1998); C.\
J.\ Hood, M.\ S.\ Chapman, T.\ W.\ Lynn, and H.\ J.\ Kimble,
Phys.\ Rev.\ Lett.\ \textbf{80}, 4157 (1998); P.\ W.\ H.\ Pinkse,
T.\ Fisher, P.\ Maunz, and G.\ Rempe, Nature \textbf{404}, 365
(2000).

\bibitem{pelton} M.\ Pelton, C.\ Santori, G.\ S.\ Solomon, O.\ Benson,
and Y.\ Yamamoto, Eur.\ Phys.\ J.\ D \textbf{18}, 179 (2002); P.\
Michler, A.\ Kiraz, C.\ Becher, L.\ Zhang, E.\ Hu, A.\ Imamoglu,
W.\ V.\ Schoenfeld, and P.\ Petroff, Phys.\ Status Solidi B
\textbf{224}, 797 (2001); J.\ McKeever,  J.\ R.\ Buck, A.\ D.\
Boozer, and H.\ J.\ Kimble, arXiv: quant-ph/0403121 (2004).

\bibitem{qipc} \textit{The Physics of Quantum Information}, edited by
D.\ Bouwmeester, A.\ Ekert, and A.\ Zeilinger (Springer-Verlag,
Berlin, 2000).

\bibitem{andersson}
\textit{Proceedings of the International Workshop on Matter Wave
Interferometry}, edited by G.\ Badurek, H.\ Rauch, and A.\
Zeilinger [Physica (Amsterdam) \textbf{151B}, No. 1-2 (1988)];
\textit{Atom Interferometry}, edited by P.\ Berman, Academic Press
(1997); A.\ Peters, K.\ Y.\ Chung, and S.\ Chu, Nature
\textbf{400}, 849 (1999); T.\ L.\ Gustavson, A.\ Landragin, and
M.\ A.\ Kasevich, J.\ Class.\ Quant.\ Grav.\ \textbf{17}, 2385
(2000) and references there in; For atomchip interferometers, see
for example:  E.\ A.\ Hinds, C.\ J.\ Vale, and M.\ G.\ Boshier,
Phys.\ Rev.\ Lett.\ \textbf{86}, 1462 (2001); W.\ H\"ansel, J.\
Reichel, P.\ Hommelhoff, and T.\ W.\ H\"ansch, arXiv:
quant-ph/0106162 (2001); E.\ Andersson, T.\ Calarco, R.\ Folman,
M.\ Andersson, B.\ Hessmo and J.\ Schmiedmayer, Phys.\ Rev.\
Lett.\ \textbf{88}, 100401 (2002).

\bibitem{computer} For a review of quantum computing in general see for
example, Phys.\ World \textbf{11} (3), 33 (1998), special issue on
quantum information, edited by M.\ B.\ Plenio and V.\ Vedral; M.\
B.\ Plenio, Contemp.\ Phys.\ \textbf{39}, 431 (1998); A.\ M.\
Steane, Rep.\ Prog.\ Phys.\ \textbf{61}, 117 (1998).

\bibitem{Kippenberg} T.\ J.\ Kippenberg, S.\ M.\ Spillane, D.\ K.\
Armani, and K.\ J.\ Vahala, Appl.\ Phys.\ Lett.\ \textbf{83}, 797
(2003); D.K.\ Armani, T.J.\ Kippenberg, S.\ M.\ Spillane, and K.\
J.\ Vahala, Nature \textbf{421}, 925 (2003); V.\ Zwiller, S.\
F\"alth, J.\ Persson, W.\ Seifert, L.\ Samuelson, and G.\ Bj\"ork,
J.\ Appl.\ Phys.\ \textbf{93}, 2307 (2003).

\bibitem{folman} R.\ Folman, P.\ Krueger, J.\ Schmiedmayer, J.\
Denschlag, and C.\ Henkel Adv.\ At.\ Mol.\ Opt.\ Phys.\
\textbf{48}, 263 (2002).

\bibitem{reichel} J.\ Reichel, Appl.\ Phys.\ B \textbf{74}, 469 (2002).

\bibitem{horak} P.\ Horak, B.\ G.\ Klappauf, A.\ Haase,
R.\ Folman, J.\ Schmiedmayer, P.\ Domokos, and E.\ A.\ Hinds,
Phys.\ Rev.\ A \textbf{67}, 043806 (2003).

\bibitem{lev} B.\ Lev, K.\ Srinivasan, P.\ Barclay, O.\ Painter,
and H.\ Mabuchi, arXiv: quant-ph/0402093 (2004).

\bibitem{lefevre} V.\ Lefevre-Seguin and S.\ Haroche, Mater.\
Sci.\ Eng.\ B \textbf{48}, 53 (1997).

\bibitem{klitzing} W.\ von Klitzing, R.\ Long, V.\ S.\ Ilchenko, J.\
Hare, and V.\ Lefevre-Seguin, New J.\ Phys.\ \textbf{3}, 14
(2001).

\bibitem{kimble} D.\ W.\ Vernooy, V.\ S.\ Ilchenko, H.\ Mabuchi,
E.\ W.\ Streed, and H.\ J.\ Kimble, Opt.\ Lett.\ \textbf{23}, 247
(1998); D.\ W.\ Vernooy, A.\ Furusawa, N.\ P.\ Georgiades, V.\ S.\
Ilchenko, and H.\ J.\ Kimble, Phys.\ Rev.\ A \textbf{57}, R2293
(1998).

\bibitem{hammer} R.\ Stoffer, K.\ R.\ Hiremath, and
M.\ Hammer, Proceedings of the International School of Quantum
Electronics, 39th course, Erice, Sicily (October 2003); M.\
Hammer, K.\ R.\ Hiremath, and R.\ Stoffer, AIP Conference
Proceedings, Melville, New York (to appear in 2004).

\bibitem{love} D.\ R.\ Rowland and J.\ D.\ Love, IEE
Proceedings-J \textbf{140}, 177 (1993).

\bibitem{rahachou} A.\ I.\ Rahachou and I.\ V.\ Zozoulenko,
arXiv: physics/0307024 (2003).

\bibitem{yamamoto} T.\ Yamamoto and M.\ Koshiba, J.\ Lightwave
Technol.\ \textbf{11}, 400 (1993).

\bibitem{hagness} S.\ C.\ Hagness, D.\ Rafizadeh, S.\ T.\ Ho, and A.\ Taflove,
J.\ Lightwave Technol.\ \textbf{15}, 2154 (1997); D.\ Rafizadeh,
J.\ P.\ Zhang, S.\ C.\ Hagness, A.\ Taflove, K.\ A.\ Stair, S.\
T.\ Ho, and R.\ C.\ Tiberio, Opt.\ Lett.\ \textbf{22}, 1244
(1997).

\bibitem{berenger} J.\ P.\ Berenger, J.\ Comput.\ Phys.\
\textbf{114}, 185 (1994); S.\ D.\ Gedney, IEEE Trans.\ Antennas
Propagation \textbf{114}, 1630 (1996); C.\ E.\ Reuter, R.\ M.\
Joseph, E.\ T.\ Thiele, D.\ S.\ Katz, and A.\ Taflove, IEEE
Microwave Guided Wave Lett.\ \textbf{4}, 344 (1994).

\bibitem{gorodetsky} M.\ L.\ Gorodetsky, A.\ A.\ Savchenkov, and V.\
S.\ Ilchenko, Opt.\ Lett.\ \textbf{21}, 453 (1996).

\bibitem{vernooy} D.\ W.\ Vernooy, V.\ S.\ Ilchenko, H.\ Mabuchi, E.\
W.\ Streed, and H.\ J.\ Kimble, Opt.\ Lett.\ \textbf{23}, 247
(1998).

\bibitem{banna} S.\ Banna, D.\ Schieber, and L.\ Sch\"{a}chter, J.\
Appl.\ Phys.\ \textbf{95}, 4415 (2004); B.\ E.\ Little and S.\ T.\
Chu, \textbf{21}, 1390 (1996); A.\ I.\ Rahachou and I.\ V.\
Zozoulenko, arXiv: physics/0305111 (2003); D.\ G.\ Hall, Opt.\
Lett.\ \textbf{6}, 601 (1981); J.\ J.\ Barroso, J.\ P.\ L.\ Neto,
and K.\ G.\ Kostov, IEEE Trans.\ Plasma Sci.\ \textbf{31}, 752
(2003).

\bibitem{lee} K.\ K.\ Lee, D.\ R.\ Lim, and L.\ C.
Kimerling, Opt.\ Lett.\ \textbf{26}, 1888 (2001); K.\ H.\ Guenter
and P.\ G.\ Wierer, Proc.\ SPIE \textbf{401}, 266 (1983).

\bibitem{javris} R.\ A.\ Javris, J.\ D.\ Love, A.\ Durandet, G.\ D.\
Conway, and R.\ W.\ Boswell, Electron.\ Lett.\ \textbf{32}, 550
(1996).

\bibitem{djordjev} K.\ Djordjev, S.-J.\ Choi, S.-J.\ Choi, and P.\
D.\ Dapkus, IEEE Photon.\ Technol.\ Lett.\ \textbf{14}, 828
(2002).

\end{thebibliography}
\end{document}